\documentclass[10pt]{article}
\usepackage[a4paper, total={6in, 8in}]{geometry}
\usepackage[usenames,dvipsnames,svgnames,table]{xcolor} 

\usepackage{algorithm}
\usepackage{algorithmic}
\floatname{algorithm}{Algorithm}

\usepackage{longtable}
\usepackage{hhline}
\usepackage{float}
\newfloat{algorithm}{tb}{lop}

\usepackage{enumitem} 
	\usepackage{rotfloat} 
	\usepackage{graphicx}
	\usepackage{epsfig}
	\usepackage{amssymb, amsmath, amsthm}
	\usepackage{tikz}
	\usetikzlibrary{3d, calc}
	\usepackage{verbatim}
	\usepackage{hyperref}
	\usepackage{natbib}
	\usepackage{authblk}
	\usepackage{multirow}
	\usepackage{subcaption} 
		\usepackage{array}
		\newcolumntype{L}{>{\centering\arraybackslash}m{0.1\linewidth}}

		\newtheorem{theorem}{Theorem}[section]
		
		\newtheorem{proposition}[theorem]{Proposition}
		\newtheorem{corollary}[theorem]{Corollary}

		\newcommand{\bfv}{\boldsymbol{v}}
		\newcommand{\bfw}{\boldsymbol{w}}
		\newcommand{\bfx}{\boldsymbol{x}}
		\newcommand{\bfy}{\boldsymbol{y}}

		\newcommand{\bbP}{\mathbb{P}}
		
		\newcommand{\bbR}{\mathbb{R}}

		\newcommand{\bbRpos}{\mathbb{R}_{+}}
		\newcommand{\bbRnonneg}{\mathbb{R}_{\geq 0}}

		\newcommand{\calM}{\mathcal{M}}
		\newcommand{\calN}{\mathcal{N}}
		\newcommand{\calO}{\mathcal{O}}

		\usepackage{xspace}
		\newcommand{\Frechet}{Fr\'{e}chet\xspace}

\begin{document}
\title{Geometric medians on product manifolds}
\author[1]{Jiewon Park}
\author[2]{Kisung You}
\affil[1]{Department of Mathematical Sciences, KAIST}
\affil[2]{Department of Mathematics, Baruch College}

\date{}
\maketitle
\begin{abstract}
	Product manifolds arise when heterogeneous geometric variables are jointly observed. While the \Frechet mean on Riemannian manifolds separates cleanly across factors, the canonical geometric median couples them, and its behavior has remained largely unexplored. In this paper, we give the first systematic treatment of this problem. After formulating the coupled objective, we establish general existence and uniqueness results that the median is unique on any Hadamard product, and remains locally unique under sharp conditions on curvature and injectivity radius even when one or more factors have positive curvature. We then prove that the estimator enjoys Lipschitz stability to perturbations and the optimal breakdown point, extending classical robustness guarantees to the product-manifold setting. Two practical solvers are proposed, including a Riemannian subgradient method with global sublinear convergence and a product-aware Weiszfeld algorithm that achieves local linear convergence. Both algorithms update the factors independently while respecting the latent coupling term, enabling implementation with standard manifold primitives. Simulations on parameter spaces of univariate and multivariate Gaussian distributions endowed with the Bures-Wasserstein geometry show that the median is more resilient to contamination than the \Frechet mean. 
\end{abstract}

\section{Introduction}

Robustness is a foundational principle in modern statistical methodology, particularly in settings where data may be contaminated by outliers, subject to model misspecification, or exhibiting heavy-tailed behavior \cite{huber_1981_RobustStatistics}. A canonical example is the sample mean, which is widely regarded as a natural estimator of central tendency under Gaussian models due to its optimality properties. However, this optimality is highly sensitive to deviations from idealized assumptions: the sample mean can exhibit substantial bias in the presence of even a single extreme observation. Such sensitivity renders the sample mean unreliable in a wide range of real-world applications.

The geometric median addresses these limitations by offering a promising alternative. Defined as the point minimizing the sum of distances to observed data, it naturally attenuates the influence of outliers. Unlike the sample mean, which aggregates squared distances and thus amplifies extremity, the geometric median prioritizes positional consensus. Beyond its robustness, it also enjoys computational simplicity and clear geometric intuition, all of which have extended its presence to curved spaces. Specifically, Riemannian manifolds now host a growing literature on statistical inference, accommodating datasets that live on spheres, rotation groups, and spaces of covariance matrices \cite{bhattacharya_2012_NonparametricInferenceManifolds, patrangenaru_2016_NonparametricStatisticsManifolds, pennec_2020_RiemannianGeometricStatistics}. For example, directional statistics leverages the geometry of the sphere \cite{mardia_2000_DirectionalStatistics}, diffusion tensor imaging works with the manifold of symmetric positive-definite matrices \cite{dryden_2009_NonEuclideanStatisticsCovariance, you_2021_RevisitingRiemannianGeometry}, and robotics operates on matrix Lie groups \cite{selig_2005_LieGroupsLie}. Foundational tools like the \Frechet mean \citep{frechet_1948_ElementsAleatoiresNature} and more recently, the Riemannian geometric median \citep{afsari_2011_Riemannian$L_p$Center, fletcher_2009_GeometricMedianRiemannian}, have been developed for such settings. These advances even extend to spaces of probability measures metrized by the Wasserstein distance \citep{you_2025_WassersteinMedianProbability}.

However, a major class of manifolds remains underexplored in this context: product manifolds. These arise when heterogeneous geometric variables are measured jointly, as in multimodal applications. For instance, combined measurements of diffusion tensors and principal directions yield data on the product of a symmetric positive-definite manifold and a sphere. In neuroimaging, functional connectivity matrices may be paired with cortical coordinates. Such product structures are ubiquitous, but robust location inference in these spaces has received little formal treatment.

A key distinction arises when comparing means and medians on product manifolds. The \Frechet mean decomposes additively across factors, allowing independent computation and analysis on each component. In contrast, the geometric median involves an $\ell_1$-like objective that couples the components through a norm structure. This non-separability introduces both theoretical and computational challenges, precluding the direct application of existing median methods designed for single manifolds.

In this paper, we develop the first comprehensive treatment of geometric medians on product manifolds, combining geometric analysis, robustness analysis, and algorithm design. Our contributions are as follows. 
\begin{enumerate}
	\item We develop a general theoretical framework for geometric medians on product manifolds, establishing existence and uniqueness results under conditions involving curvature bounds and injectivity radii of the component manifolds.
	\item We derive robustness guarantees in the form of perturbation bounds and breakdown properties, showing that the geometric median retains desirable stability characteristics under mild geometric conditions.
	\item We introduce two algorithmic strategies for computing the geometric median based on subgradient descent and the Weiszfeld algorithm. Both algorithms operate component-wise while incorporating the coupling structure of the objective. Convergence properties are established under suitable regularity assumptions.
\end{enumerate}

The rest of the paper is organized as follows. Section~\ref{sec:preliminaries} reviews the geometry of product manifolds and the formulation of geometric medians in the Riemannian setting. In Section~\ref{sec:theory}, the main theoretical results on existence, uniqueness, and robustness are presented. Section~\ref{sec:computation} introduces computational algorithms and analyzes their convergence properties. Section~\ref{sec:examples} illustrates the methodology through representative examples. We conclude by discussing open directions for future studies in Section \ref{sec:conclusion}.

\section{Preliminaries}\label{sec:preliminaries}

We start by reviewing the mathematical foundations for the study of geometric medians on product manifolds at the minimal level. We first recall the definition and properties of the geometric median in the Riemannian setting, followed by a description of product manifold geometry, which plays a key role in both our theoretical and computational developments.

\subsection{Geometric medians on Riemannian manifolds}

Let $(\calM, g)$ be a complete Riemannian manifold with geodesic distance function $d:\calM\times \calM\to \bbRnonneg$. Given a random sample $x_1, \ldots, x_n \in \calM$, along with weights $w_1 ,\ldots, w_n \in \bbR$ such that $\sum_{i=1}^n w_i = 1$ and $w_i > 0$ for all $i\in [n]:=\lbrace 1, \ldots, n\rbrace$, the Riemannian $L_p$ center of mass \citep{afsari_2011_Riemannian$L_p$Center} for $p\geq 1$ is defined as
\begin{equation}\label{def:lp_center_of_mass}
	p\text{-center}(x_1, \ldots, x_n) := \underset{x\in\calM}{\arg\min}~\sum_{i=1}^n w_i d(x,x_i)^p.
\end{equation}
This formulation extends the $L_p$ center of mass in Euclidean spaces to general Riemannian manifolds. For instance, setting $\calM = \bbR^d$ and $p=2$ induces the solution of \eqref{def:lp_center_of_mass} as the standard weighted average. For general manifolds, the problem and its solution are known as the \Frechet mean when $p=2$, which is a direct generalization of the sample mean. Another important case is when $p=1$, the minimizer of whcih has been known as the geometric median  \citep{fletcher_2009_GeometricMedianRiemannian}.

Denote $F(x) = \sum_{i=1}^n w_i d(x, x_i)$ the objective function in the geometric median problem. When the point $x\in \calM$ is such that the geodesic distance between $x$ and each datum $x_i$ is unique and length-minimizing, $F(x)$ is directionally differentiable and locally Lipschitz. At such points, the subdifferential of $F$ contains a vector 
\begin{equation*}
	-\sum_{i=1}^n w_i \cdot \frac{\log_x (x_i)}{\|\log_x (x_i)\|} \in \partial F(x),
\end{equation*}
where $\log_x (x_i) \in T_x \calM$ denotes the Riemannian logarithmic map that pushes $x_i \in \calM$ to the tangent space at $x$, giving the direction and magnitude of the geodesic from $x$ to $x_i$. This subgradient formulation generalizes the Euclidean subdifferential of the $\ell_1$ norm to the Riemannian setting \citep{ferreira_1998_SubgradientAlgorithmRiemannian}. When $x=x_j$ for some $j \in [n]$, or when multiple minimizing geodesics exist, the subdifferential is a set and includes a convex set of descent directions. 

Existence of the geometric median is guaranteed under mild conditions, the weakest of which is completeness of $\calM$. However, uniqueness is much more subtle, contingent on both the curvature of the manifold and the dispersion of the data distribution. In particular, on Hadamard manifolds that are complete, simply-connected, and nonpositively curved, the uniqueness is immediate by the convexity of the distance function. For positively curved manifolds, uniqueness may fail unless the data lie within a convex geodesic ball of sufficiently small radius \citep{bhattacharya_2012_NonparametricInferenceManifolds}.

\subsection{Geometry of product manifolds}

Let $(M,g_M)$ and $(N,g_N)$ be smooth, connected Riemannian manifolds. The product manifold $\calM = M \times N$ naturally inherits a Riemannian structure given by the product metric $g_\calM := g_M \oplus g_N$ \citep{lee_1997_RiemannianManifoldsIntroduction}. For any point $(p,q) \in \calM$, the tangent space decomposes as a direct sum
\begin{equation}\label{def:product_tangent}
	T_{(p,q)}(M\times N) = T_p M \oplus T_q N,
\end{equation}
and the inner product between tangent vectors $(\bfv_1,\bfw_1), (\bfv_2,\bfw_2) \in T_{(p,q)}\calM$ is defined by 
\begin{equation}\label{def:product_metric}
	g_\calM ((\bfv_1, \bfw_1), (\bfv_2,\bfw_2)) := g_M (\bfv_1,\bfv_2) + g_N (\bfw_1,\bfw_2).
\end{equation}
For the rest of this paper, we will call $M$ and $N$ as factor manifolds, or simply factors, to denote components in defining a product manifold.

Geodesics in $\calM$ are given by pointwise pairing of geodesics from each factor. That is, if $\gamma_M:[0,1] \to M$ and $\gamma_N:[0,1]\to N$ are geodesics in $M$ and $N$ respectively, then $\gamma(t) = (\gamma_M(t), \gamma_N(t))$ defines a geodesic in $\calM$. The exponential map on $\calM$ satisfies
\begin{equation*}
	\exp_{(p,q)}(\bfv,\bfw) = (\exp_p \bfv, \exp_q \bfw),
\end{equation*}
and the logarithmic map similarly decomposes as 
\begin{equation*}
	\log_{(p,q)}(p',q') = (\log_p p', \log_q q'),
\end{equation*}
for $(\bfv,\bfw) \in T_{(p,q)}\calM$ and $(p',q') \in \calM$. The geodesic distance on $\calM$ is induced by the product metric 
\begin{equation*}
	d_\calM((p_1,q_1), (p_2,q_2)) = \sqrt{d_M(p_1,p_2)^2 + d_N(q_1,q_2)^2},
\end{equation*}
where $d_M$ and $d_N$ denote the geodesic distances in $M$ and $N$, respectively.

We close this section by discussing the distinctive nature of product manifolds. The decomposable structure simplifies many geometric computations and plays a crucial role in both the theoretical and algorithmic treatment of estimators on product manifolds. However, the global geometric behavior of $\calM$ intricately depends on the curvature properties of $M$ and $N$. For instance, if both factors are complete and has nonpositive curvature, their product inherits similar geometric regularity. That is, it becomes a Hadamard manifold which guarantees on global geodesic convexity and uniqueness of certain minimizers. On the other hand, it one or both have positive curvature, such global guarantees may no longer hold and analysis must be restricted with a sufficiently small geodesic ball to control convexity conditions. These subtleties play a crucial role in the behavior of the geometric median on product manifolds as we will examine later.

\section{Theory}\label{sec:theory}

\subsection{Problem formulation}

Let $\calM = M\times N$ be a product manifold equipped with the product Riemannian metric $g_\calM$ and the distance function $d_\calM$ as defined in Section \ref{sec:preliminaries}. Suppose we have a coupled random sample $\lbrace (x_i, y_i)\rbrace_{i=1}^n \subset \calM$, with associated positive weights $w_1, \ldots, w_n > 0$ such that $\sum_{i=1}^n w_i = 1$. The geometric median is defined as a minimizer $(p^*,q^*) \in \calM$ of the objective function
\begin{equation}\label{def:objective_median} 
	F_{\text{median}}(p,q) = \sum_{i=1}^n w_i d_\calM((p,q), (x_i,y_i)) 
	=\sum_{i=1}^n w_i  \sqrt{d_M(p,x_i)^2 + d_N(q,y_i)^2}.
\end{equation}
This formulation naturally couples the variables $p\in M$ and $q\in N$ on factor manifolds through the norm structure of the product distance. For comparison, consider the \Frechet mean objective 
\begin{equation*}
	F_{\text{mean}}(p,q) = 
	\sum_{i=1}^n w_i d_\calM((p,q), (x_i,y_i))^2 =
	\sum_{i=1}^n w_i  d_M(p,x_i)^2 + \sum_{i=1}^n w_i d_N(q,y_i)^2,
\end{equation*}
which admits separation of the objective into additive components, allowing independent optimization over $M$ and $N$. Unfortunately, the geometric median objective $F_{\text{median}}$ lacks such separability. Hence, the minimizer $(p^*,q^*)$ cannot be obtained by solving two independent problems and this interdependency entails nontrivial implications for both theoretical properties and computational treatment.

\subsection{Existence}

We first establishing the existence of a geometric median on the product manifold $\calM$, the argument of which relies on classical results in variational analysis.

\begin{theorem}\label{theory:existence}
	Let $M$ and $N$ be complete Riemannian manifolds and let $\calM = M\times N$ denote the associated product manifold with the product metric. Given a random sample $\lbrace (x_i,y_i)\rbrace_{i=1}^n$ and weights $w_1,\ldots,w_n > 0$ satisfying $\sum_{i=1}^n w_i = 1$, 
	there exists at least one minimizer of the objective function \begin{equation*}
		F_{\text{median}}(p,q) = \sum_{i=1}^n w_i d_\calM((p,q), (x_i,y_i)).
	\end{equation*}
	That is, the geometric median on $\calM$ exists.
\end{theorem}
\begin{proof}
	Each map $(p,q)\mapsto d_\calM((p,q), (x_i,y_i))=\sqrt{d_M(p,x_i)^2+d_N(q,y_i)^2}$ is continuous per the properties of the Riemannian distance function \citep{lee_2012_IntroductionSmoothManifolds} and its composition, hence so is $F_{\mathrm{median}}$. 
	
	By completeness of $M$ and $N$, its product $\calM$ endowed with the $\ell_2$ product distance is complete \citep{lee_2018_IntroductionRiemannianManifolds}. By the Hopf-Rinow theorem, $\calM$ is proper, i.e., closed and bounded sets are compact \citep{docarmo_1992_RiemannianGeometry}. If $(p_k,q_k)$ leaves every compact subset of $\calM$, then for any fixed $i$ either $d_M(p_k,x_i)\to\infty$ or $d_N(q_k,y_i)\to\infty$, hence $d_\calM((p_k,q_k),(x_i,y_i))\to\infty$ and thus $F_{\mathrm{median}}(p_k,q_k)\to\infty$. Therefore $F_{\mathrm{median}}$ is coercive, its sublevel sets are compact, and by the direct method it attains a minimum.
\end{proof}

We note that neither $M$ nor $N$ is assumed to be compact. If so, the coercivity of the objective function is readily available since all continuous functions on compact Riemannian manifolds are bounded and attain their extrema. Therefore, the existence of a geometric median is guaranteed even when only one of the two component manifolds is noncompact.

\subsection{Uniqueness}

We now study conditions for uniqueness of the geometric median on a product manifold. Throughout, we assume the sample $S=\{(x_i,y_i)\}_{i=1}^n\subset\mathcal{M}$ is geodesically non-collinear, i.e., it is not contained in the image of any geodesic segment or equivalently, there do not exist $z\in\mathcal{M}$, $v\in T_z\mathcal{M}$, and an interval $I$ with $S\subset\{\exp_z(tv):t\in I\}$). In Hadamard spaces this yields strict geodesic convexity of the sum of distances and hence global uniqueness. When positive curvature is present, we work inside a strongly convex (normal) ball $B((p_0,q_0),r)$, the radius of which will be specified, so that minimizing geodesics and logarithmic maps are unique on the region and the same strict-convexity argument gives uniqueness within the ball.

\subsubsection{Curvature structure of product manifolds}

The uniqueness of the geometric median is closely tied to the curvature properties of the underlying manifold. We begin by investigating the sectional curvature of product manifolds in terms of those of their factors.

\begin{proposition}\label{theory:curvature_bound}
Let $M$ and $N$ be Riemannian manifold manifolds with $\text{sec}_M \leq \kappa_M$ and $\text{sec}_N \leq \kappa_N$, respectively, for $\kappa_M, \kappa_N \geq 0$. Then the sectional curvature of the product manifold $M\times N$ equipped with the product metric is bounded above by $\max(\kappa_M, \kappa_N)$. 
\end{proposition}
\begin{proof}
For linearly independent tangent vectors $\bfx = (\bfx_1,\bfx_2)$ and $\bfy = (\bfy_1,\bfy_2)$ in $T_{(p,q)}(M\times N)$, define $$|\bfx \wedge \bfy |^2_{M\times N}:= |\bfx|^2_{M\times N} |\bfy|^2_{M\times N} - \langle \bfx,\bfy \rangle^2_{M\times N}$$ as in for instance \citep{lee_2012_IntroductionSmoothManifolds}. Then $|\bfx \wedge \bfy |^2_{M\times N} \ne 0$. We observe that

\begin{equation}\label{proof_proposition_equation1}
	\begin{aligned}
		|\bfx\wedge \bfy|_{M\times N}^2 &= |\bfx|^2_{M\times N} |\bfy|^2_{M\times N} - \langle \bfx,\bfy \rangle^2_{M\times N} \\ 
		&= (|\bfx_1|^2_M + |\bfx_2|^2_M)(|\bfy_1|^2_N + |\bfy_2|^2_N) - (\langle \bfx_1, \bfy_1\rangle_M + \langle \bfx_2,\bfy_2\rangle_N)^2 \\
		&=|\bfx_1\wedge \bfy_1|_M^2 + |\bfx_2 \wedge \bfy_2|_N^2 + |\bfx_1|_M^2 |\bfy_2|_N^2 + |\bfx_2|_N^2 |\bfy_1|_M^2 - 2\langle \bfx_1, \bfy_1\rangle_M \langle \bfx_2,\bfy_2\rangle_N.
	\end{aligned}
\end{equation}
For brevity, denote $(|\bfx_1|_M, |\bfx_2|_N, |\bfy_1|_M, |\bfy_2|_N)=(a,b,c,d)$, where each quantity is nonnegative. Dividing both sides of  \eqref{proof_proposition_equation1} by $|\bfx\wedge \bfy|^2_{M\times N}$ leads to 
\begin{equation}\label{proof_proposition_equation2}
	1 = \frac{|\bfx_1\wedge \bfy_1|_M^2}{|\bfx\wedge \bfy|^2_{M\times N}} + \frac{|\bfx_2 \wedge \bfy_2|_N^2}{|\bfx\wedge \bfy|^2_{M\times N}} + 
	\frac{a^2d^2 + b^2 c^2 - 2\langle \bfx_1, \bfy_1\rangle_M \langle \bfx_2,\bfy_2\rangle_N}{|\bfx\wedge \bfy|^2_{M\times N}}.
\end{equation}
We now examine the last term of \eqref{proof_proposition_equation2}. By the Cauchy-Schwarz inequality,
\begin{align*}
	a^2 d^2 + b^2 c^2 - 2\langle \bfx_1, \bfy_1\rangle_M \langle \bfx_2,\bfy_2\rangle_N \geq a^2d^2 + b^2c^2 - 2abcd = (ad-bc)^2 \geq 0.
\end{align*}
Therefore, we conclude that 
\begin{equation}\label{proof_proposition_equation3}
	1 \geq \frac{|\bfx_1\wedge \bfy_1|_M^2}{|\bfx\wedge \bfy|^2_{M\times N}} + \frac{|\bfx_2 \wedge \bfy_2|_N^2}{|\bfx\wedge \bfy|^2_{M\times N}},
\end{equation}
which will be used to control the weights in a convex combination of curvature terms in what follows.

Recall that the sectional curvature of $M\times N$ of the 2-plane spanned by $\bfx, \bfy$ is given by 
\begin{equation*}
	\text{sec}_{M\times N}(\bfx\wedge \bfy) = \frac{\langle R_{M\times N}(\bfx,\bfy) \bfy, \bfx \rangle}{|\bfx\wedge \bfy|^2_{M\times N}},
\end{equation*}
where $R_{M\times N}$ denotes the Riemannian curvature tensor of $M\times N$. Since any mixed 2-plane is flat,
\begin{equation*}
	\langle R_{M\times N}(\bfx,\bfy) \bfy, \bfx \rangle = \langle R_{M}(\bfx_1,\bfy_1) \bfy_1, \bfx_1 \rangle + 
	\langle R_{N}(\bfx_2,\bfy_2) \bfy_2, \bfx_2 \rangle.
\end{equation*}
Now since $\text{sec}_M \leq \kappa_M$ and $\text{sec}_N \leq \kappa_N$, by the definition of sectional curvature we obtain
\begin{align*}
	\langle R_{M}(\bfx_1,\bfy_1) \bfy_1, \bfx_1 \rangle &\leq \kappa_M |\bfx_1\wedge \bfy_1|^2_M,\\
	\langle R_{N}(\bfx_2,\bfy_2) \bfy_2, \bfx_2 \rangle &\leq \kappa_N |\bfx_2\wedge \bfy_2|^2_N.
\end{align*}
Hence, we obtain
\begin{align*}
	\sec_{M\times N}(\bfx\wedge \bfy) &\leq \frac{\kappa_M |\bfx_1\wedge \bfy_1|_M^2 + \kappa_N |\bfx_2 \wedge \bfy_2|^2}{|\bfx\wedge \bfy|_{M\times N}^2} \\ 
	&\leq \frac{\max(\kappa_M, \kappa_N)\cdot  ( |\bfx_1\wedge \bfy_1|_M^2 +  |\bfx_2 \wedge \bfy_2|^2)}{|\bfx\wedge \bfy|_{M\times N}^2} \\
	&\leq \frac{\max(\kappa_M, \kappa_N)\cdot |\bfx\wedge \bfy|_{M\times N}^2}{|\bfx\wedge \bfy|_{M\times N}^2} = \max(\kappa_M, \kappa_N),
\end{align*}
where the last inequality uses  \eqref{proof_proposition_equation3}. This completes the proof.
\end{proof}

This proposition provides a geometric characterization of the curvature profile of product manifolds in that the curvature of any 2-plane in $\calM$ is bounded from above by the largest curvature among the factors.

\subsubsection{Global uniqueness in Hadamard product manifolds}

A direct consequence of Proposition~\ref{theory:curvature_bound} is the following uniqueness statement for products of manifolds with nonpositive curvature.

\begin{proposition}\label{theory:result1_hadamard}
Let $M$ and $N$ be Hadamard manifolds; that is, complete simply connected Riemannian manifolds with nonpositive sectional curvature. Then the product $\calM=M\times N$ is also Hadamard, and the geometric median of any non-collinear finitely many weighted points in $\calM$ is unique. 
\end{proposition}
\begin{proof}
Since both $M$ and $N$ are Hadamard, their curvatures satisfy  $\text{sec}_M \leq 0$ and $\text{sec}_N \leq 0$, and  each factor is complete and simply connected. From Proposition \ref{theory:curvature_bound}, the product $\calM = M\times N$ has nonpositive section curvature. Furthermore, the product of two simply connected manifolds is also simply connected. Hence, $\calM$ is Hadamard. On a Hadamard manifold, the distance function $x\mapsto d(x,x_i)$ is convex, an in particular strictly convex along a geodesic not passing through $x_i$. Since the data points $x_i$ are not collinear, a positively weighted sum of the distance functions $d(x_i, \cdot)$ is strictly convex. Therefore, the geometric median which is the minimizer of this functional is unique \citep{rockafellar_1997_ConvexAnalysis}.
\end{proof} 

This uniqueness simplifies both the theoretical analysis and algorithmic aspect of the geometric median on products of Hadamard manifolds, including common examples such as the Euclidean space, the hyperbolic space, and the space of symmetric positive-definite matrices.

\subsubsection{Local uniqueness for positive curvature}

In the last subsection we examined Hadamard manifolds. However, many important applications involve manifolds with bounded positive curvature, compact components, or mixtures of such structures. In these settings, model spaces frequently encountered in practice do not admit global uniqueness of the geometric median, thereby necessitating a local analysis. We present a unified result that ensures uniqueness of the geometric median on product manifolds under mild constraints. This framework also encompasses mixed-curvature settings where factors include Euclidean spaces and the unit hypersphere, both of which are common in directional statistics. To this end, we establish sufficient conditions for the local uniqueness of the geometric median based on curvature upper bounds and injectivity radii.

\begin{theorem}\label{theory:result2_local_uniqueness}
Let $M$ and $N$ be complete Riemannian manifolds with bounded sectional curvatures $\text{sec}_M \leq\kappa_M, \text{sec}_N \leq \kappa_N$ where $\kappa_M, \kappa_N \ge 0$. Let $\kappa:=\max(\kappa_M, \kappa_N)$. Suppose there exists a point $(p_0, q_0) \in \calM$ such that a random sample $\lbrace (x_i, y_i)\rbrace_{i=1}^n \subset \calM$ lies within the geodesic ball $B((p_0,q_0), r) \subset \calM$ with 
\begin{equation*}
	r < \min \left\lbrace 
	\textrm{inj}_M (p_0), \textrm{inj}_N (q_0), \frac{\pi}{4\sqrt{\kappa}}
	\right\rbrace,
\end{equation*}
where $\textrm{inj}_M (x)$ is the injectivity radius of a manifold $M$ at $x\in M$. Then the geometric median uniquely exists within the geodesic ball $B((p_0, q_0), r)$.
\end{theorem}
\begin{proof}
We start by establishing the convexity of the squared distance under curvature bounds. It is a standard result in comparison geometry that the squared distance function $z \mapsto d(z,z_0)^2$ is convex along geodesics within a ball of radius less than $\pi/(2\sqrt{\kappa})$ centered at a point $z_0$, and strictly convex if the ball lies outside the cut locus of $z_0$ \citep{petersen_2006_RiemannianGeometry, afsari_2011_Riemannian$L_p$Center}. Since the product manifold $\calM$ inherits an upper curvature bound $\kappa := \max(\kappa_M, \kappa_N)$ by Proposition \ref{theory:curvature_bound}, for each $(x_i, y_i)\in \calM$, the map 
\begin{equation*}
	(p,q) \mapsto d_{\calM}((p,q), (x_i, y_i))^2 = d_M (p,x_i)^2 + d_N(q,y_i)^2
\end{equation*}
is strictly convex on the ball $B((p_0, q_0), r)$ once the radius $r$ is smaller than the injectivity radius at each factor and less than the comparison threshold $\pi/(2\sqrt{\kappa})$ \citep{cheeger_2008_ComparisonTheoremsRiemannian}. 

Since $B((p_0,q_0),r)\subset B((x_i,y_i),\rho)$ for every $i$ with $2r<\rho\le \pi/(2\sqrt{\kappa})$, the map $(p,q)\mapsto d_{\calM}((p,q),(x_i,y_i))$ is geodesically convex on $B((p_0,q_0),r)$ strictly convex unless the sample is collinear. Hence, the positive weighted sum $F_{\mathrm{median}}$ is strictly geodesically convex on this ball, admitting a unique minimizer within this ball \citep{afsari_2011_Riemannian$L_p$Center}.
\end{proof}

Theorem \ref{theory:result2_local_uniqueness} provides a general criterion for the local uniqueness of the geometric median. The radius condition reflects a standard trade-off between positive curvature and geodesic convexity. Suppose $\kappa = 0$ as in flat or Hadamard spaces, the uniqueness holds globally. When $\kappa > 0$, the uniqueness holds within sufficiently small open sets. This generalizes convexity-based uniqueness results for the \Frechet mean to the non-smooth geometric median setting \citep{afsari_2011_Riemannian$L_p$Center}.

A particularly useful implication arises when one factor is  nonpositively curved while the other is compact with bounded positive curvature, such as the unit hypersphere. In such mixed-curvature settings, the uniqueness still holds with the radius constraint only on the positively curved factor as formalized in the following corollary. Since the argument is similar we omit the proof.

\begin{corollary}\label{theory:result3_corollary_of_result2}
Suppose $M$ is a Hadamard manifold and $N$ is compact with $\text{sec}_N \leq \kappa_N$ for $\kappa_N > 0$.  Then, for any sufficiently small ball $B((p_0, q_0), r) \subset \calM$ containing a random sample   $\lbrace (x_i, y_i)\rbrace_{i=1}^n \subset \calM$ with radius 
\begin{equation*}
	r < \min \left\lbrace 
	\textrm{inj}_N (q_0), \frac{\pi}{4\sqrt{\kappa_N}}
	\right\rbrace,
\end{equation*}   
the geometric median is unique and lies within $B((p_0, q_0), r)$.
\end{corollary}

\subsection{Robustness}

In this subsection, we establish theoretical guarantees regarding the robustness of the geometric median on product manifolds. To recall, robustness refers to the stability of the estimator under small perturbations of the data  and its resilience to outliers \citep{huber_1981_RobustStatistics}. These properties are well known in the Euclidean setting and have been studied in various geometric contexts, including Riemannian manifolds  \citep{fletcher_2009_GeometricMedianRiemannian}. We show that similar guarantees hold in the product setting, both locally and globally, under suitable conditions.

We begin by formalizing the stability of the geometric median under data perturbations. The result assumes that both the original and perturbed datasets lie within a convex geodesic ball whose radius is determined by curvature and injectivity radius as in Theorem \ref{theory:result2_local_uniqueness}.

\begin{proposition}\label{theory:robustness_local_robustness}
Assume the setting of Theorem~\ref{theory:result2_local_uniqueness}, so that $F(p,q)=\sum_i w_i\,d_\mathcal{M}((p,q),(x_i,y_i))$ is uniformly geodesically convex on $B((p_0,q_0),r)$ and has a unique minimizer $m=(p^*,q^*)$ in this ball. Then there exists $\mu>0$ which is the local strong geodesic convexity modulus of $F$ on the ball, depending on $\kappa$, $r$, and the configuration, to the following effect. Let $\{(x_i',y_i')\}$ be a perturbed sample in the same ball with minimizer $m'=(p^{\prime *},q^{\prime *})$, and set
\[
\varepsilon:=\sum_{i=1}^n w_i\,d_\mathcal{M} \left((x_i,y_i),(x_i',y_i')\right).
\]
Then
\[
d_\mathcal{M}(m',m)^2 \le \frac{4}{\mu}\,\varepsilon
\le \frac{4}{\mu}\sum_{i=1}^n w_i\,\big(\varepsilon_i^M+\varepsilon_i^N\big),
\]
where $d_M(x_i,x_i')\le \varepsilon_i^M$ and $d_N(y_i,y_i')\le \varepsilon_i^N$.
\end{proposition}

\begin{proof}
Denote $F'(p,q)=\sum_i w_i\,d_\mathcal{M}((p,q),(x_i',y_i'))$. By the triangle inequality, for every $(p,q)$ we have
\[
\bigl|F(p,q)-F'(p,q)\bigr|
=\Bigl|\sum_i w_i\bigl(d((p,q),(x_i,y_i))-d((p,q),(x_i',y_i'))\bigr)\Bigr|
\le \sum_i w_i\, d\bigl((x_i,y_i),(x_i',y_i')\bigr)
= \varepsilon.
\]
In the product metric, $d_\calM((x_i,y_i),(x_i',y_i'))=\sqrt{d_M(x_i,x_i')^2+d_N(y_i,y_i')^2}
\le d_M(x_i,x_i')+d_N(y_i,y_i')$. Hence, 
\[
\varepsilon \le \sum_{i=1}^n w_i\sqrt{(\varepsilon_i^M)^2+(\varepsilon_i^N)^2}
\le \sum_{i=1}^n w_i\bigl(\varepsilon_i^M+\varepsilon_i^N\bigr).
\]

By Theorem~\ref{theory:result2_local_uniqueness}, on the strongly convex ball 
$B((p_0,q_0),r)$ the function $F$ is (strictly) geodesically convex, smooth away from the data, and  has positive definite Riemannian Hessian at $m$ under geodesic non-collinearity. By compactness of the ball and continuity of the Hessian, there exists $\mu>0$ such that $F(z)\ge F(m)+\frac{\mu}{2}\,d(z,m)^2$ for all $z \in B((p_0,q_0),r)$.

Let $m'=(p^{\prime *},q^{\prime *})$ be the minimizer of $F'$ in the ball. Then, 
\[
F(m') \le F'(m')+\varepsilon = \min F' + \varepsilon 
\le F'(m)+\varepsilon \le F(m)+2\varepsilon.
\]
On the other hand, 
\[
F(m') \ge F(m) + \frac{\mu}{2}\, d(m',m)^2.
\]
Combining the two  gives
\[
\frac{\mu}{2}\, d(m',m)^2 \le 2\varepsilon
\quad\Longrightarrow\quad
d(m',m)^2 \le \frac{4}{\mu}\,\varepsilon.
\]
Substituting the bound on $\varepsilon$ from above yields the stated estimate.
\end{proof}


Proposition~\ref{theory:robustness_local_robustness} gives a local stability result. On a strongly convex ball where $F$ enjoys quadratic growth with modulus $\mu>0$, the median map is $1/2$–H\"{o}lder with respect to the weighted average perturbation size. Thus small local perturbations of the inputs produce small displacements of the median, with a constant depending only on the geometry and the configuration through $\mu$.

Next, we characterize the estimator's resilience to adversarial contamination. The following result establishes that the breakdown point of the geometric median on product manifolds is asymptotically optimal.

\begin{proposition}\label{theory:robustness_breakdown}
Let $\calM=M\times N$ be a noncompact finite-dimensional Riemannian manifold. Given weighted data $z_i = (x_i, y_i) \in \calM$ for $i=1, \ldots, n$ with weights $w_i>0$, $\sum_i w_i=1$, denote by $W_I:=\sum_{i\in\cal I}w_i$ the total weight of a subset $\cal I$ of $\{1, \ldots, n\}$ to be contaminated. 
Let $\tilde z$ denote any geometric median of the contaminated sample $\{z'_i\}$ with $z'_i=z_i$ for $i\notin\cal I$ and arbitrary $z'_i\in\calM$ for $i\in\cal I$.

(a) If $W_I<1/2$, then the geometric median is bounded by the clean data, no matter how $\{z'_i\}$ is contaminated.

(b) If $W_I>1/2$, then there exist contaminations $\{z'_i\}$ for which every geometric median can be made arbitrarily far from the clean cloud, hence unbounded as the contaminated points are sent to infinity. 

(c) If $W_I = 1/2$, there exist contaminations for which the set of geometric medians is unbounded and some medians can be taken arbitrarily far, although not all medians need diverge.

Consequently, the finite-sample breakdown point of the geomtric median equals $1/2$.
\end{proposition}
\begin{proof} 
Let $F(z) = \sum_{i=1}^n w_i d_\calM(z, z_i)$ and $F'(z) = \sum_{i=1}^n w_i d_\calM (z, z_i')$ and denote by $J := \lbrace 1, \ldots, n\rbrace\backslash I$ the clean indices with $W_J = 1-W_I$. 

Let $K$ be a compact subset of $\mathcal M$ containing the clean points $z_j$, $j\in J$. By the triangle inequality and the fact that $d(z,z_j)\ge d(z,K)$ we observe that
\[F'(z)=\sum_{j \in J}w_j d(z,z_j)+\sum_{i\in I}w_id(z,z_i') \ge W_J \cdot d(z,K)+\sum_{i\in I}w_i(d(z_i',K)-d(z,K))=(1-2W_I)d(z,K)+\sum_{i\in I}w_i d(z_i',K).\]

First consider the case $W_I<1/2$, or equivalently, $1-2W_I>0$. Then $\inf_{z\in K}F'(z)\ge F'(\bar z)\ge(1-2W_I)d(\bar z,K)+\sum_{i\in I}w_i d(z_i',K) \ge (1-2W_I)d(\bar z, K)$. Therefore, $d(\bar z,K) \le \frac{1}{1-2W_I}\left(\inf_{z\in K}F'(z)-\sum_{i\in I}w_i d(z_i',K)\right)$. Taking any $z_0 \in K$ and denoting by $diam(K)$ the diameter of $K$, which is the maximum distance between two points in $K$, observe that by the triangle inequality,
\begin{align*}
	\inf_{z\in K}F'(z)-\sum_{i\in I}w_i d(z_i',K) &\le F'(z_0)-\sum_{i\in I}w_i d(z_i',K) \\
	&=\sum_{j\in J} w_jd(z_0, z_j)+\sum_{i\in I}\left(d(z_i',z_0)-d(z_i',K)\right)\\
	&\le W_J\cdot diam(K)+W_I \cdot diam(K),
\end{align*}
hence  \[d(\bar z, K) \le \frac{diam(K)}{1-2W_I}.\] 

Now consider the case $W_I>1/2$. Let us take our contamination to be such that $z_i'=z'$ tends to infinity. We observe that
\[F'(z) \ge \sum_{i\in I}w_i d(z_i',K)-(2W_I-1)d(z,K)=W_Id(z',K)-(2W_I-1)d(z,K).\]
In particular, $F'(\bar z) \ge W_Id(z',K)-(2W_I-1)d(\bar z,K)$.
On the other hand observe that
\[F'(\bar z)\le F'(z')=\sum_{j\in J}w_jd(z',z_j)\le W_J\cdot d(z',K)+W_J\cdot diam(K).\]
Combining we have that \[W_Id(z',K)-(2W_I-1)d(\bar z,K) \leq W_J\cdot d(z',K)+W_J\cdot diam(K)=(1-W_I)\cdot d(z',K)+W_J\cdot diam(K),\]
or equivalently,
\[d(\bar z,K) \geq d(z',K)-\frac{W_J}{2W_I-1}\cdot diam(K).\]
Therefore, if $d(z',K) \to \infty$, then $d(\bar z, K) \to \infty$. In other words for this choice of contamination the geometric median becomes unbounded.

Lastly when $W_I = 1/2$, set $z_i' = z'$ for $i\in I$ and send $d(z',K)\to \infty$. Recall the inequality already established:
\begin{equation*}
	F'(z) \ge (1-2W_I) d(z,K) + \sum_{i\in I} w_i d(z_i', K) = \frac{1}{2} d(z',K),
\end{equation*}
for all $z \in \calM$. Evaluating at $z'$ and using $d(z', z_j) \le d(z', K) + diam(K)$ gives 
\begin{equation*}
	F'(z') \le \frac{1}{2} d(z',K) + \frac{1}{2}diam(K)\quad \Longrightarrow \quad \underset{z}{\inf F'(z)} \in 
	\left[
	\frac{1}{2}d(z',K), \frac{1}{2}d(z',K) + \frac{1}{2}diam(K)
	\right].
\end{equation*}
Particularly, as $d(z',K)\to \infty$, the set of minimizers is unbounded as there exist geometric medians at arbitrarily large distance from $K$. However, unlike the $W_I > 1/2$ case, not every median must diverge, e.g., the one-dimensional example with half the mass at 0 and half at $l\to \infty$. This completes the proof.
\end{proof}

This establishes a breakdown point of $1/2$. That is, as long as the total contaminated weight is below $1/2$, the geometric median cannot be driven arbitrarily far, whereas above $1/2$, an adversary can force divergence. Combined with the local quadratic-growth stability bound from Proposition~\ref{theory:robustness_local_robustness}, this recovers the classical robustness guarantees of the geometric median in the product-manifold setting, in line with the single-manifold case \citep{fletcher_2009_GeometricMedianRiemannian}.

\section{Computation}\label{sec:computation}

The non-separability of the median objective function $F_{\text{median}}$ necessitates modification of the standard apparatus. In this section, we present computational strategies for estimating the geometric median on product manifolds. Due to the nonsmooth nature of $F_{\text{median}}$, we first formulate a general subgradient-based approach and show how the classical Riemannian Weiszfeld algorithm arises as a special case. We then study the convergence behavior of these algorithms under mild assumptions on the geometry of the manifold and the regularity of the data. 

\subsection{Algorithms}

The product structure of $\calM = M\times N$ allows computation of the geometric median in a component-wise manner, with updates independently performed on each factor manifold with an interleaving term that connects the two. That is, the optimization problem on the product manifold naturally reduces to a coupled problem over $M$ and $N$, where each update step involves operations in the respective tangent space $T_p M$ and $T_q N$. This structure invites direct application of manifold optimization routines without requiring complex constructions on the full product space. We posit that a random sample $\lbrace (x_i,y_i)\rbrace_{i=1}^n \subset \calM$ is given with positive weights $\lbrace w_i\rbrace_{i=1}^n$ that sums to 1. For simplicity, we denote the objective $F_{\text{median}}$ simply as $F$ throughout the rest of this section.

\subsubsection{Riemannian subgradient algorithm}
At any point $(p,q) \in \calM$, the subdifferential of $F$ decomposes naturally as 
\begin{equation*}
\partial F(p,q) \subset \partial F_p (p,q) \times \partial F_q(p,q),
\end{equation*}
where $\partial_p F(p,q) \subset T_p M$ and $\partial_q F(p,q) \subset T_q N$ are the partial subdifferentials of $F$ with respect to each component. Using the standard notation, we can express partial differentials explicitly as 
\begin{align*}
\partial_p F(p,q) &~\reflectbox{$\in$} - \sum_{i: p\neq x_i} w_i \cdot \frac{\log_p (x_i)}{d_\calM ((p,q), (x_i, y_i))} + \sum_{i:p=x_i} w_i \cdot B_p,\\
\partial_q F(p,q) &~\reflectbox{$\in$} - \sum_{j: q\neq y_j} w_j \cdot \frac{\log_q (y_j)}{d_\calM ((p,q), (x_j, y_j))} + \sum_{j:q=y_j} w_j \cdot B_q,
\end{align*}
where $\log_p (x):M\to T_p M$ and $\log_q (y):N\to T_q N$ are logarithmic maps \citep{absil_2008_OptimizationAlgorithmsMatrix}, and $B_p = \lbrace v  \in T_pM \mid \|v\| \leq 1\rbrace$ and 
$B_q = \lbrace w  \in T_q N \mid \|w\| \leq 1\rbrace$ are the closed unit balls in the respective tangent spaces and  represent  the set-valued contributions from points where the distance function is non-differentiable. Note that the summation is in the sense of Minkowski.

For each $i$, set $a_i:=d_M(p,x_i)$, $b_i:=d_N(q,y_i)$ and $d_i:=\sqrt{a_i^2+b_i^2}$, and define
$F_i(p,q):=w_i\,d_i$. The contribution of the $i$-th term to the partial subdifferentials can be categorized as follows. When both factors admit the regular differentiability with $a_i>0$ and $b_i > 0$, we have 
\[
\partial_p F_i(p,q)=\left\lbrace-\,w_i\,\frac{\log_p(x_i)}{d_i}\right\rbrace,
\qquad
\partial_q F_i(p,q)=\left\lbrace -\,w_i\,\frac{\log_q(y_i)}{d_i}\right\rbrace.
\]
Next, consider when the equality holds for one of the factor manifolds. Without loss of generality, consider $p=x_i$ and $q \neq y_i$ or equivalently,  $a_i = 0$ and $b_i \neq 0$. Then,
\[
F_i(p,q)=w_i\sqrt{\,\|u\|^2+b_i^2\,}\quad\text{with }u:=\log_p(x_i),
\]
so the map $u\mapsto w_i\sqrt{\|u\|^2+b_i^2}$ is smooth at $u=0$ and has gradient $0$.
Therefore,
\[
\partial_p F_i(p,q)=\{0\},\qquad
\partial_q F_i(p,q)=\left\lbrace -\,w_i\,\frac{\log_q(y_i)}{b_i}\right\rbrace.
\]
The last scenario is when $a_i=0$ and $b_i=0$, i.e., $(p,q)=(x_i,y_i)$, with maximal nondifferentiability. Then the joint subdifferential is the closed unit ball scaled by $w_i$ in the product tangent space,
$
\partial F_i(p,q)=w_i\,\overline{B}_{T_{(p,q)}\mathcal{M}},
$
so in particular
\[
\partial_p F_i(p,q)\supset w_i\,B_p,\qquad
\partial_q F_i(p,q)\supset w_i\,B_q,
\]
where $B_p\subset T_pM$ and $B_q\subset T_qN$ are the closed unit balls.

Putting the components together, we obtain a subgradient descent scheme on $\calM$, where updates are computed independently along each factor. Let $(\boldsymbol{\xi}_p^{(k)}, \boldsymbol{\xi}_q^{(k)}) \in \partial F(p^{(k)}, q^{(k)})$ denote a valid choice of subgradients at the $k$-th iteration. The iterates are updated according to the following rules 
\begin{equation*}
p^{(k+1)} = \exp_{p^{(k)}}(-\eta_k \cdot \boldsymbol{\xi}_p^{(k)})\quad\text{and}\quad 
q^{(k+1)} = \exp_{q^{(k)}}(-\eta_k \cdot \boldsymbol{\xi}_q^{(k)}),
\end{equation*}
where $\exp_p(v):T_pM \to M$ and $\exp_q(w):T_q N \to N$ are exponential maps on each factor and $\eta_k$ is a scalar-valued step size.  This formulation updates coordinates at the same time while two updates involve an interleaving term $d_\calM ((p^{(k)},q^{(k)}), (x_i, y_i))$ that respects the formulation of the geometric median. In practice, a specific choice of subgradient can be made by selecting a representative from each set-valued term, such as the element of minimal norm in $\partial_p F(p,q)$ and $\partial_q F(p,q)$ or even random vectors of small magnitude. We note that any measurable selection guarantees validity of descent direction upon updates under standard conditions.

\subsubsection{Weiszfeld algorithm}

A well-known alternative to subgradient descent is a fixed-point iteration that enforces the standard first-order condition of vanishing gradients. When the current iterate $(p,q)$ is far away from all of $(x_i, y_i)$'s, the subdifferentials $\partial_p F(p,q)$ and $\partial_q F(p,q)$ are singletons and the conditions on vanishing gradients become
\begin{equation*}
\sum_{i=1}^n w_i \cdot \frac{\log_p (x_i)}{d_\calM((p,q), (x_i, y_i))} = 0, \qquad \sum_{i=1}^n w_i \cdot \frac{\log_q (y_i)}{d_\calM((p,q), (x_i, y_i))} = 0.
\end{equation*}
Solving for a square root of these expressions motivates a fixed-point approach where the update is derived by exponentiating average of logarithmic directions weighted by the inverse of the distances. The resulting scheme, known as the Riemannian Weiszfeld algorithm \citep{fletcher_2009_GeometricMedianRiemannian}, is derived as follows:
\begin{equation*}
p^{(k+1)} = \exp_{p^{(k)}}\left(\sum_{i=1}^n \tilde{w}_i \log_{p^{(k)}}(x_i)\right)
\quad\text{and}\quad 
q^{(k+1)} = \exp_{q^{(k)}}\left(\sum_{i=1}^n \tilde{w}_i \log_{q^{(k)}}(y_i)\right),
\end{equation*}
with adaptive weights $\tilde{w}_i^{(k)}$'s that are given by 
\begin{equation*}
\tilde{w}_i^{(k)} = \left(\sum_{j=1}^n 
\frac{w_j}{d_\calM((p^{(k)}, q^{(k)}), (x_j, y_j))}
\right)^{-1}\cdot \frac{w_i}{d_\calM((p^{(k)}, q^{(k)}), (x_i, y_i))}.
\end{equation*}
Each update corresponds to a weighted \Frechet mean on Riemannian manifolds where the contribution of each data point is scaled inversely by its distance from the current iterate. This yields an implicit normalization of the descent direction and avoids explicit tuning of step size.

The Weiszfeld algorithm is parameter-free and often exhibits linear convergence near the solution under appropriate regularity. It is particularly effective when the geometric median lies deep within a geodesically convex neighborhood of the data. However, its fixed-point nature and reliance on inverse-distance scaling can result in instability near singularities, requiring care in implementation. There are several ad hoc remedies to cope with such scenarios \cite{beck_2015_WeiszfeldsMethodOld}. One is to regularize terms with small distances $d_\calM$ by replacing it with or adding a small constant $\varepsilon$. Another strategy is to replace each update with a convex combination of the current iterate and the raw Weiszfeld direction, the process called damping. There are also algorithmic remedies to restart the algorithm, switch to a subgradient update, or even terminate at the point. 

In contrast to subgradient descent, the Weiszfeld method exploits structure but lacks robustness in degenerate regimes. In practice, hybrid schemes that use Weiszfeld iterations when far from singularities and fall back to subgradient updates when near data points can be particularly effective \cite{beck_2015_WeiszfeldsMethodOld}.

\subsection{Convergence analysis}

We now analyze the convergence behavior of the two algorithms that were previously introduced. While the objective $F_{\text{median}}$ is nonsmooth and nonconvex in general, both the subgradient descent and Weiszfeld algorithms exhibit provable convergence under appropriate geometric constraints.

First, we investigate convergence behavior of the Riemannian subgradient algorithm. In the special case where both factors $M$ and $N$ are Hadamard manifolds, we know that the function $F_{\text{median}}$ is geodesically convex under the non-collinearity condition, i.e., for any geometric $\gamma:[0,1] \to \calM$, the function $t \mapsto F_{\text{median}}(\gamma(t))$ is convex. In this setting, the subgradient method enjoys global convergence guarantees from nonsmooth Riemannian optimization theory \citep{zhang_2016_FirstorderMethodsGeodesically}. Specifically, if subgradients are uniformly bounded and the step sizes satisfy $\eta_k = \eta_0/\sqrt{k+1}$ for some $\eta_0 > 0$, then the following holds:
\begin{equation*}
\underset{0\leq j\leq k}{\min}~ F_{\text{median}}(p^{(j)}, q^{(j)}) - F_\text{median}(p^*, q^*) = \mathcal{O}\left(\frac{\log k}{\sqrt{k}}\right), 
\end{equation*}
where $(p^*, q^*)$ are minimizers of the objective.

If $M$ or $N$ has positive curvature, global convexity is no longer available, yet convexity can still be locally recovered within a sufficiently small geodesic ball. That is, if the data and all iterates remain within such ball, the same convergence guarantees still hold locally.

\begin{theorem}\label{theory:convergence_subgradient}
Let $\calM=M\times N$ be a product of complete Riemannian manifolds with curvature upper bounds $\text{sec}_M \leq \kappa_M$ and 
$\text{sec}_N \leq \kappa_N$ with  $\kappa := \max(\kappa_M, \kappa_N)$, where $\kappa_M, \kappa_N \ge 0$. Suppose a random sample $\lbrace (x_i, y_i)\rbrace_{i=1}^n$ and the initial point $(p^{(0)}, q^{(0)})$ all lie within a geodesic ball centered at $(p_0,q_0)$ with a radius $r$ as prescribed in Theorem \ref{theory:result2_local_uniqueness}. Assume that the iterates $(p^{(k)}, q^{(k)})$ along the path of the subgradient method remain in the ball and the step size is chosen as $\eta_k = \eta_0/\sqrt{k+1}$ for some $\eta_0 > 0$. Then, the geometric median is unique in the ball $B((p_0, q_0), r)$ and it achieves a sublinear rate of convergence
\begin{equation*}
	\underset{0\leq j\leq k}{\min}~ F_{\text{median}}(p^{(j)}, q^{(j)}) - F_\text{median}(p^*, q^*) = \mathcal{O}\left(\frac{\log k}{\sqrt{k}}\right).
\end{equation*}
\end{theorem}
\begin{proof}
The subgradient \((\boldsymbol{\xi}_p, \boldsymbol{\xi}_q) \in T_p M \times T_q N\) at any \((p, q) \in \calM\) satisfies
\[
\boldsymbol{\xi}_p := -2 \sum_{i: p \ne x_i} w_i \cdot \frac{\log_p(x_i)}{d_\calM((p, q), (x_i, y_i))}, \qquad
\boldsymbol{\xi}_q := -2 \sum_{i: q \ne y_i} w_i \cdot \frac{\log_q(y_i)}{d_\calM((p, q), (x_i, y_i))}.
\]
Since the iterates and all data points are assumed to lie in the compact geodesic ball \(B((p_0, q_0), r)\), both the numerator and denominator terms in the above expression are bounded. Hence, the norm of the full subgradient vector \(\boldsymbol{\xi}^{(k)} := (\boldsymbol{\xi}_p^{(k)}, \boldsymbol{\xi}_q^{(k)})\) is uniformly bounded. That is, $\|\boldsymbol{\xi}^{(k)}\| \le G < \infty$ for all $k$.

We now apply the standard convergence result for Riemannian subgradient descent on geodesically convex functions \citep{zhang_2016_FirstorderMethodsGeodesically}. Since the objective is geodesically convex on a convex geodesic ball, the subgradients are uniformly bounded, and the step size is chosen as \(\eta_k = \eta_0 / \sqrt{k+1}\), 
it follows that
\[
\min_{0 \le j \le k} \left[ F(p^{(j)}, q^{(j)}) - F(p^*, q^*) \right] \le \frac{D^2 + \eta_0^2 G^2 \log(k+1)}{2 \eta_0 \sqrt{k+1}},
\]
where \(D := d_\calM((p^{(0)}, q^{(0)}), (p^*, q^*))\) is the Riemannian distance to the unique minimizer \((p^*, q^*)\). This establishes the desired sublinear convergence rate of  $\calO(\log k/\sqrt{k})$.
\end{proof}
The convergence rate $\mathcal{O}\left(\frac{\log k}{\sqrt{k}}\right)$ stated in Theorem \ref{theory:convergence_subgradient} arises from applying a specific bound using the step size sequence $\eta_k = \eta_0/\sqrt{k+1}$. Although this bound shows a factor $\log k$, the standard optimal convergence rate for the decrease in the value of the function of the subgradient method in geodesically convex and Lipschitz functions is known to be $\mathcal{O}\left(\frac{1}{\sqrt{k}}\right)$. This tighter rate can typically be achieved with alternative step-size selection strategies, such as fixing a priori the total number of iterations or using step sizes, e.g., Polyak, that depend on the norm of the subgradient or through a more refined convergence analysis.

We now turn to the Weiszfeld algorithm adapted to our setting. When the manifold $\calM$ is Hadamard, the algorithm enjoys global convergence to the unique geometric median, provided the iterates remain distinct from all data points. This is due to the fact that the Weiszfeld update corresponds to a normalized fixed-point iteration of the subgradient condition, which is well-defined and contractive in convex geodesic regions.

In positively curved or mixed-curvature settings, however, convergence is no longer global. Nonetheless, under the same local convexity condition as Theorem \ref{theory:convergence_subgradient}, the Weiszfeld algorithm converges to the unique minimizer in a sufficiently small geodesic ball, provided that the iterates avoid the singularities induced by the data points.

\begin{corollary}\label{theory:convergence_weiszfeld}
Under the same assumptions as in Theorem \ref{theory:convergence_subgradient}, suppose the iterates of the Riemannian Weiszfeld algorithm remain within the ball $B((p_0, q_0), r)$ and that no iterates collapse onto any data point $(x_i,y_i)$ for all $i\in[n]$. Then, the algorithm converges to the unique geometric median in the ball. Moreover, if the distances $d_\calM((p^{(k)},q^{(k)}), (x_i,y_i))$ are uniformly bounded away from zero, the convergence is locally linear.
\end{corollary}
\begin{proof}
Since the assumptions match those in Theorem~\ref{theory:convergence_subgradient}, the objective function \(F_{\text{median}}\) is geodesically convex within the ball \(B((p_0, q_0), r)\), and admits a unique minimizer \((p^*, q^*)\).

Each Weiszfeld update is defined as a retraction along the direction
\[
\boldsymbol{\Delta}_p^{(k)} := 
\left(\sum_{i=1}^n \frac{w_i}{d_\calM((p^{(k)}, q^{(k)}), (x_i, y_i))}\right)^{-1}\cdot 
\sum_{i=1}^n w_i \cdot \frac{\log_{p^{(k)}}(x_i)}{d_\calM((p^{(k)}, q^{(k)}), (x_i, y_i))}, \]
and similarly for \(\boldsymbol{\Delta}_q^{(k)}\), where all terms are well-defined given that $((p^{(k)}, q^{(k)}) \neq (x_i, y_i)$. The fact that the iterates remain in the ball of radius $r$ ensures that both the log maps and distance terms vary smoothly. Therefore, the update map defines a continuous self-map on a compact convex domain. By standard contraction arguments \citep{fletcher_2009_GeometricMedianRiemannian},  the Weiszfeld iteration converges to the unique fixed point when the iterates are well-separated from singularities.

In addition, suppose all pairwise distances \(d_\calM((p^{(k)}, q^{(k)}), (x_i, y_i))\) are bounded below by a fixed \(\delta > 0\). Then the  weights in the denominator remain uniformly bounded and the iteration map is a Lipschitz continuous map with Lipschitz constant strictly less than 1 in a neighborhood of the solution \citep{vardi_2000_MultivariateL1medianAssociated}. This guarantees local linear convergence to the minimizer.
\end{proof}

This result confirms that the Weiszfeld algorithm is both theoretically sound and computationally attractive in product manifold settings, particularly when the geometry permits local convexity. The assumption that iterates remain distinct from the data points excludes a measure-zero singular set where the denominator in the update becomes ill-defined. In practice, this is enforced by the aforementioned strategies, including damping or regularization.

The local linear convergence rate hinges on the conditioning of the problem near the minimizer. Specifically, when the geometric median lies well-separated from the data, the denominator terms in the update remain uniformly bounded away from zero, and the algorithm behaves like a contractive fixed-point map. This is analogous to strong convexity in Euclidean settings, though here it arises from local convexity and smoothness of the Riemannian distance function within the ball. In contrast, when data points cluster tightly or lie near the median, the conditioning deteriorates, and convergence may slow down or stall due to near-singular behavior. This sensitivity motivates ad hoc hybrid strategies. For instance, one may begin with robust subgradient iterations that are insensitive to non-differentiability and switch to Weiszfeld updates once the iterates approach the minimizer. Such hybrid schemes could help to balance global robustness with local acceleration and are particularly effective when the median lies near high-curvature regions of the manifold.

\section{Examples}\label{sec:examples}
In this section, we illustrate the proposed framework for computing geometric medians on product manifolds using simulated datasets. Specifically, we focus on the space of univariate and multivariate Gaussian distributions, where each distribution is viewed as a point on a product manifold endowed with the Bures-Wasserstein geometry  \citep{takatsu_2011_WassersteinGeometryGaussian}. These examples highlight the estimator’s robustness to contamination and the feasibility of the proposed algorithms.

First, consider the space of univariate Gaussian laws  $\calN(\mu,\sigma^2)$, parametrized by a mean $\mu \in \bbR$ and a standard deviation $\sigma > 0$. This space can be identified with $\bbR \times \bbRpos$, equipped with the 2-Wasserstein distance. In this case, the distance between two Gaussians $\bbP_1 = \calN(\mu_1,\sigma_1^2)$ and $\bbP_2 = \calN(\mu_2,\sigma_2^2)$ admits a closed-form expression:
\begin{equation}\label{def:bures_univariate}
d_W (\bbP_1, \bbP_2)^2 = (\mu_1 - \mu_2)^2 + (\sigma_1 - \sigma_2)^2.
\end{equation}
While this geometry appears Euclidean, it reflects optimal transport geometry and forms a special case of the broader Bures-Wasserstein structure.

We simulate $n = 1000$ observations $\calN(\mu_i, \sigma_i^2)$ by drawing $\mu_i \sim \calN(-1, 1/4)$ and $\sigma_i^2 \sim \text{Beta}(5,5)$, forming the signal distribution. The most probable realization is centered at $\calN(-1, 1/2)$, corresponding to the modal values of the generative distributions for the parameters. These parameter pairs are treated as points on $\calM = \bbR \times \bbR_{+}$, with distance measured according to \eqref{def:bures_univariate}.

In order to assess robustness, we introduce contamination by replacing a fraction $\alpha$ of the data with outliers. Specifically, $\lfloor \alpha n \rfloor$ samples are replaced with Gaussians whose parameters are drawn from $\mu_i \sim \calN(5,1)$ and $\sigma_i^2 \sim 5 \cdot \text{Beta}(5,5)$. We compute both the \Frechet mean, which corresponds to the Wasserstein barycenter in the literature of optimal transport, and the geometric median of the parameterized points. The former is available in closed form, while the latter is computed using the Weiszfeld algorithm.

\begin{figure}[ht]
\centering
\begin{subfigure}[b]{0.30\textwidth}
	\includegraphics[width=\textwidth]{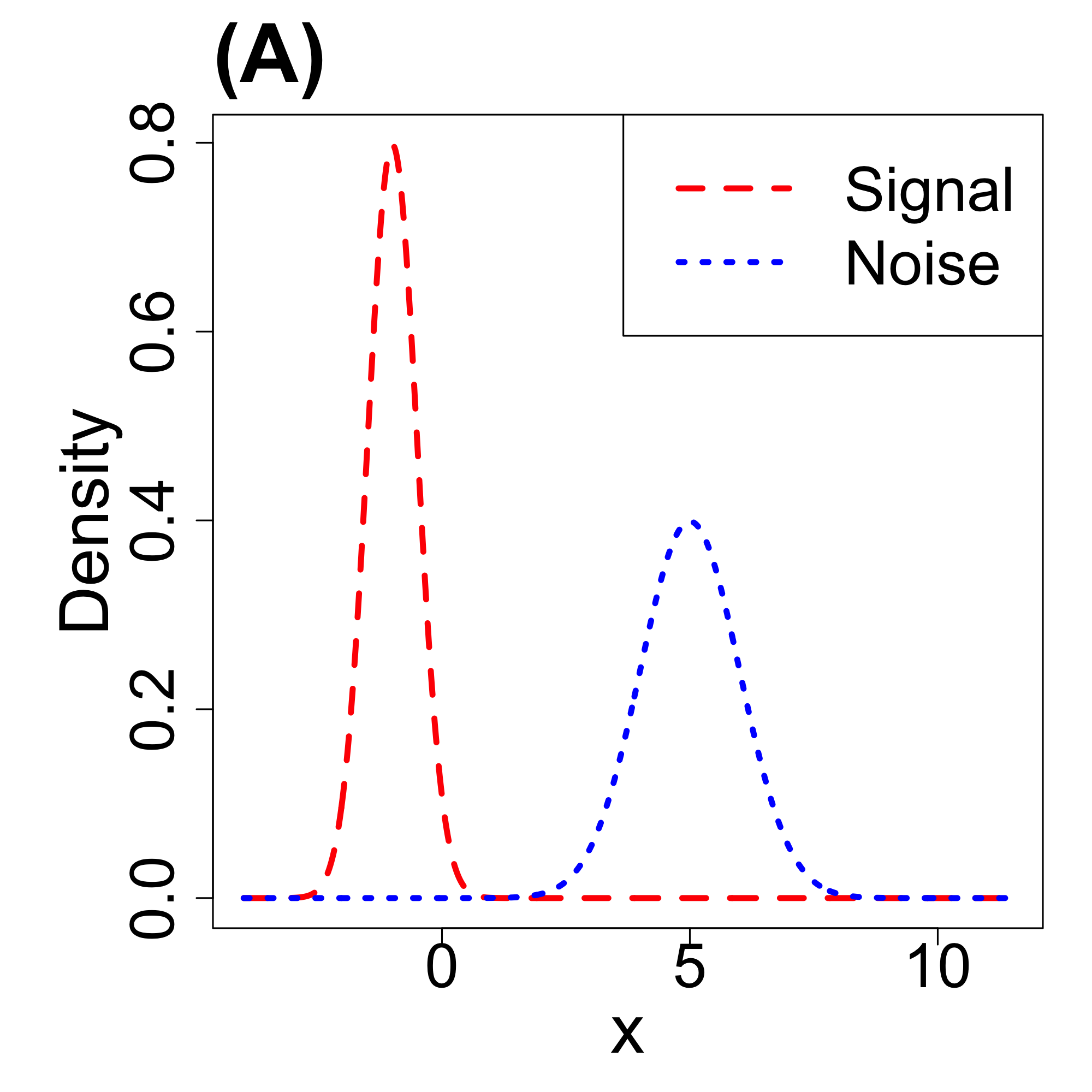}
\end{subfigure}
\hfill
\begin{subfigure}[b]{0.30\textwidth}
	\includegraphics[width=\textwidth]{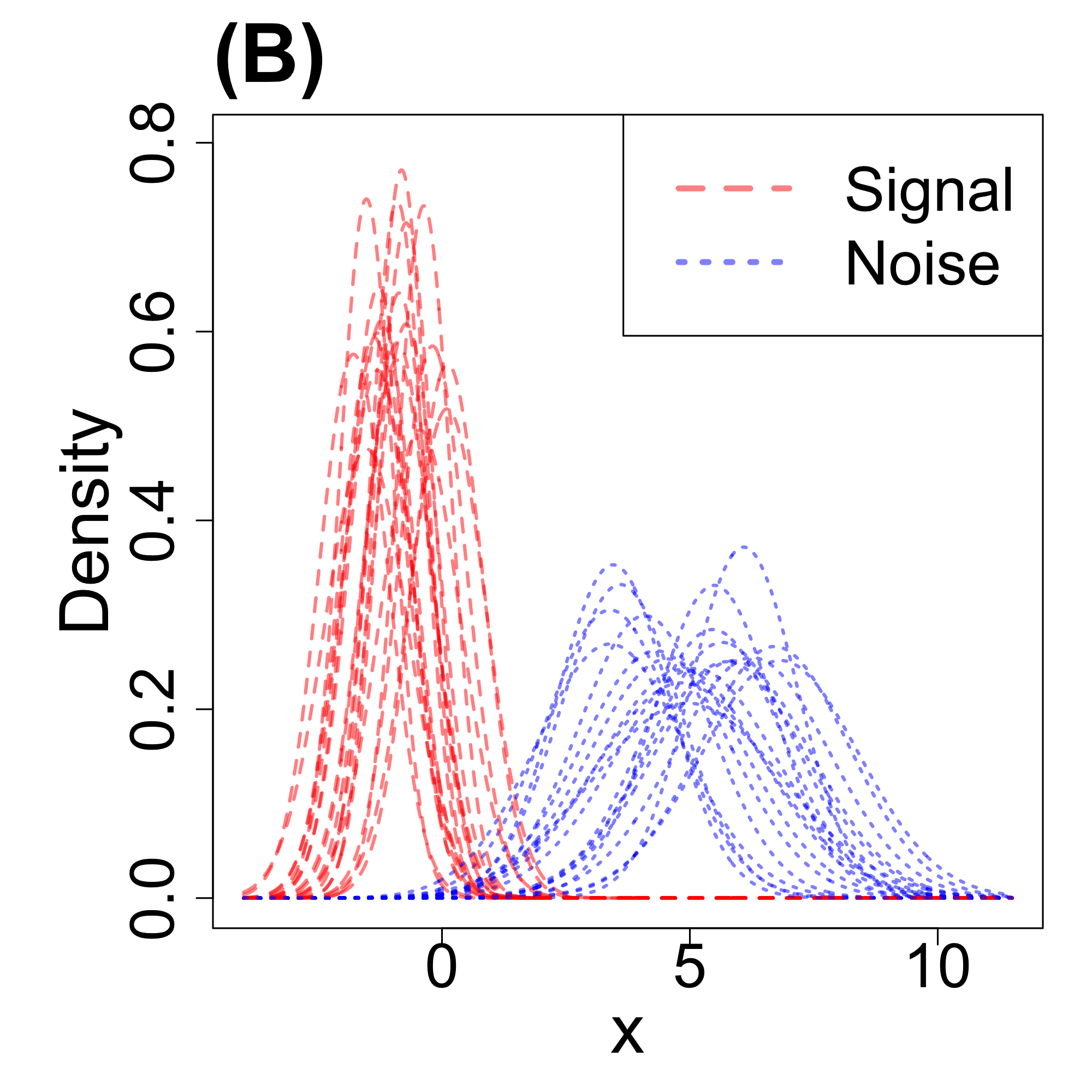}
\end{subfigure}
\hfill
\begin{subfigure}[b]{0.30\textwidth}
	\includegraphics[width=\textwidth]{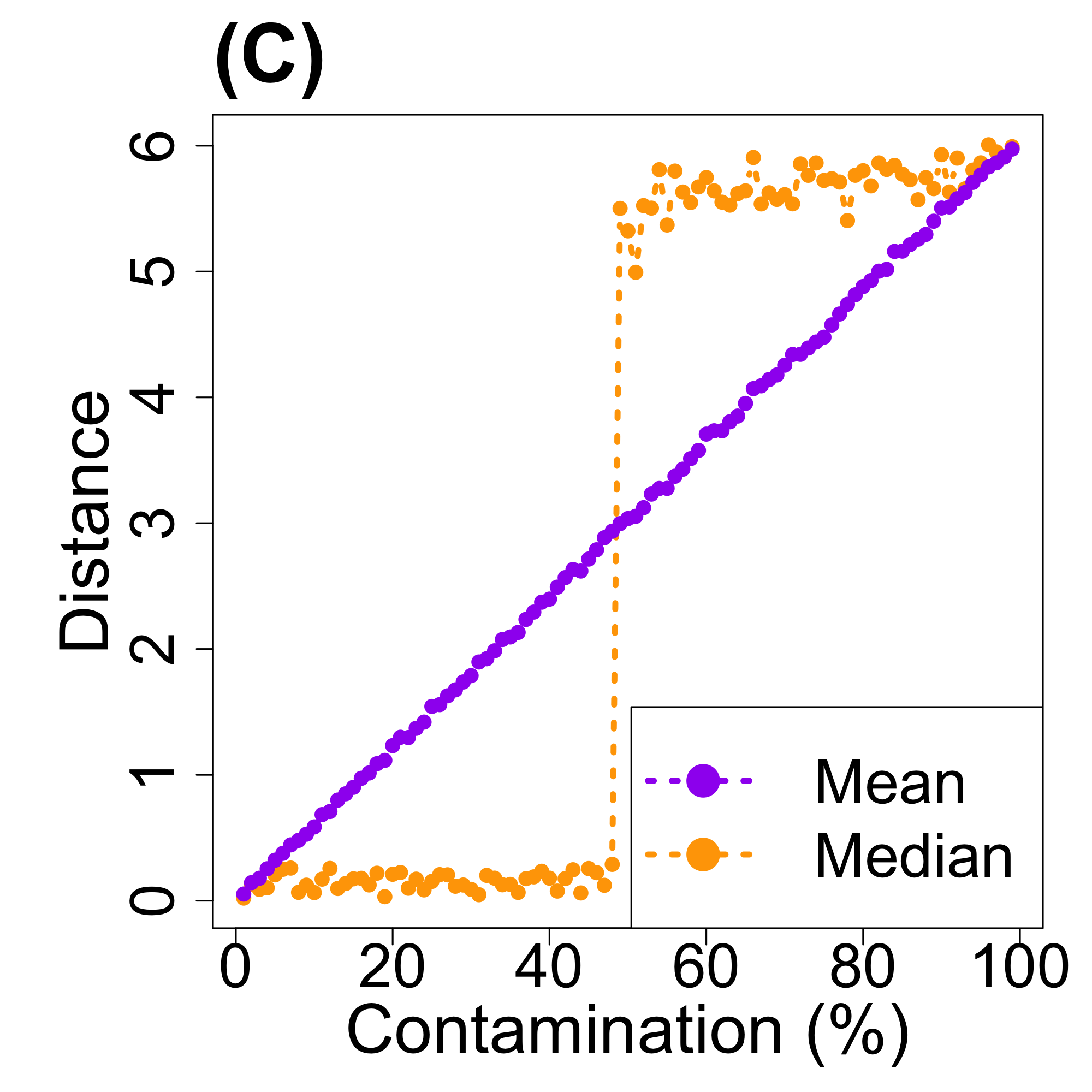}
\end{subfigure}
\caption{Visualization of the univariate Gaussian example. (A) Model densities of the signal and noise are presented. (B) A representative set of realized signal and noise distributions is shown. (C) Estimation error as a function of contamination rate is given for both \Frechet mean and geometric median.}
\label{fig:example1}
\end{figure}

Fig. \ref{fig:example1} illustrates the results. As the contamination rate increases, the discrepancy between the \Frechet mean and the model signal grows linearly. In contrast, the geometric median remains stable until the contamination level approaches 50\%, at which point a sharp transition in performance is observed, reflecting the aforementioned theoretical breakdown point. Beyond this threshold, the geometric median shifts toward the model noise, indicating that the notion of ``signal'' itself may become ill-posed when it constitutes a minority of the data.

We now extend the experiment to multivariate Gaussian distributions $\calN_d(\mu,\Sigma)$ characterized by mean vector $\mu \in \bbR^d$ and covariance matrix $\Sigma \in \bbR^{d \times d}$, where $\Sigma$ is symmetric and positive-definite. For $\bbP_1 = \calN_d(\mu_1, \Sigma_1)$ and $\bbP_2 = \calN_d(\mu_2, \Sigma_2)$, the 2-Wasserstein distance is
\begin{equation}\label{def:bures_multivariate}
d_W (\bbP_1, \bbP_2)^2 = \|\mu_1 - \mu_2\|^2 + \text{trace}\left(\Sigma_1 + \Sigma_2 - 2(\Sigma_2^{1/2}\Sigma_1\Sigma_2^{1/2})^{1/2}\right),
\end{equation}
where $\Sigma^{1/2}$ is the matrix square root, i.e., $\Sigma^{1/2}\cdot \Sigma^{1/2} = \Sigma$. While this expression simplifies to  \eqref{def:bures_univariate} when $d=1$, the space of multivariate Gaussian distributions is different from the univariate case in several key senses. First, while the 2-Wasserstein distance in one dimension decomposes cleanly into additive contributions from mean and variance differences, the multivariate case involves intricate interactions between covariance matrices, requiring matrix square roots and trace terms that reflect both scale and orientation. Second, the geometry of the space becomes non-Euclidean as the set of multivariate Gaussians endowed with the 2-Wasserstein metric forms a Riemannian manifold, where geodesics and interpolation paths are curved and shape-aware, unlike the linear interpolation in $\bbR$. Finally, optimal transport maps in higher dimensions are no longer monotonic functions but instead involve linear transformations that align mass in both direction and spread, introducing substantial geometric and computational complexity.

We employ the previous experimental design by simulating  $n=1000$ replicates of the signal distribution $\calN_d(\mu_i, \Sigma_i)$, each of which is a perturbed version of $\calN_d(0, I_d)$, where $I_d$ is the $d \times d$ identity matrix. Instead of directly sampling the parameters, each replicate is generated by drawing a sample of size $2d$ from $\calN_d(0, I_d)$ and computing the corresponding maximum likelihood estimates (MLEs), which serve as the parameters $(\mu_i, \Sigma_i)$ of the simulated signal distributions. To model contamination, we replace $\lfloor \alpha n \rfloor$ distributions with noise. The noise distributions are sampled from $\calN_d(10, \Sigma_{\text{AR}})$, where $\Sigma_{\text{AR}}$ is the autoregressive covariance matrix of an AR(1) process, defined as
\begin{equation*}
\Sigma(i,j) = \rho^{|i-j|}, \quad 1 \leq i,j \leq d,
\end{equation*}
for a decay parameter $\rho \in (0,1)$, as described in \cite{bickel_2008_RegularizedEstimationLarge}. As with the signal generation, each noise distribution is based on a sample of size $2d$, and its MLEs provide the parameters of the noise distributions under the Gaussian model. We consider varying the dimensionality $d \in \lbrace 10, 50, 100 \rbrace$ and the decay parameter $\rho \in \lbrace 0.1, 0.5, 0.9 \rbrace$, and compute both the \Frechet mean and the geometric median in the parameter space. In this experiment, we restrict the contamination level to at most 49\%, in order to avoid the regime where the proportion of noise exceeds that of the signal, as discussed in the previous example.
\begin{figure}[ht]
\centering
\includegraphics[width=.9\linewidth]{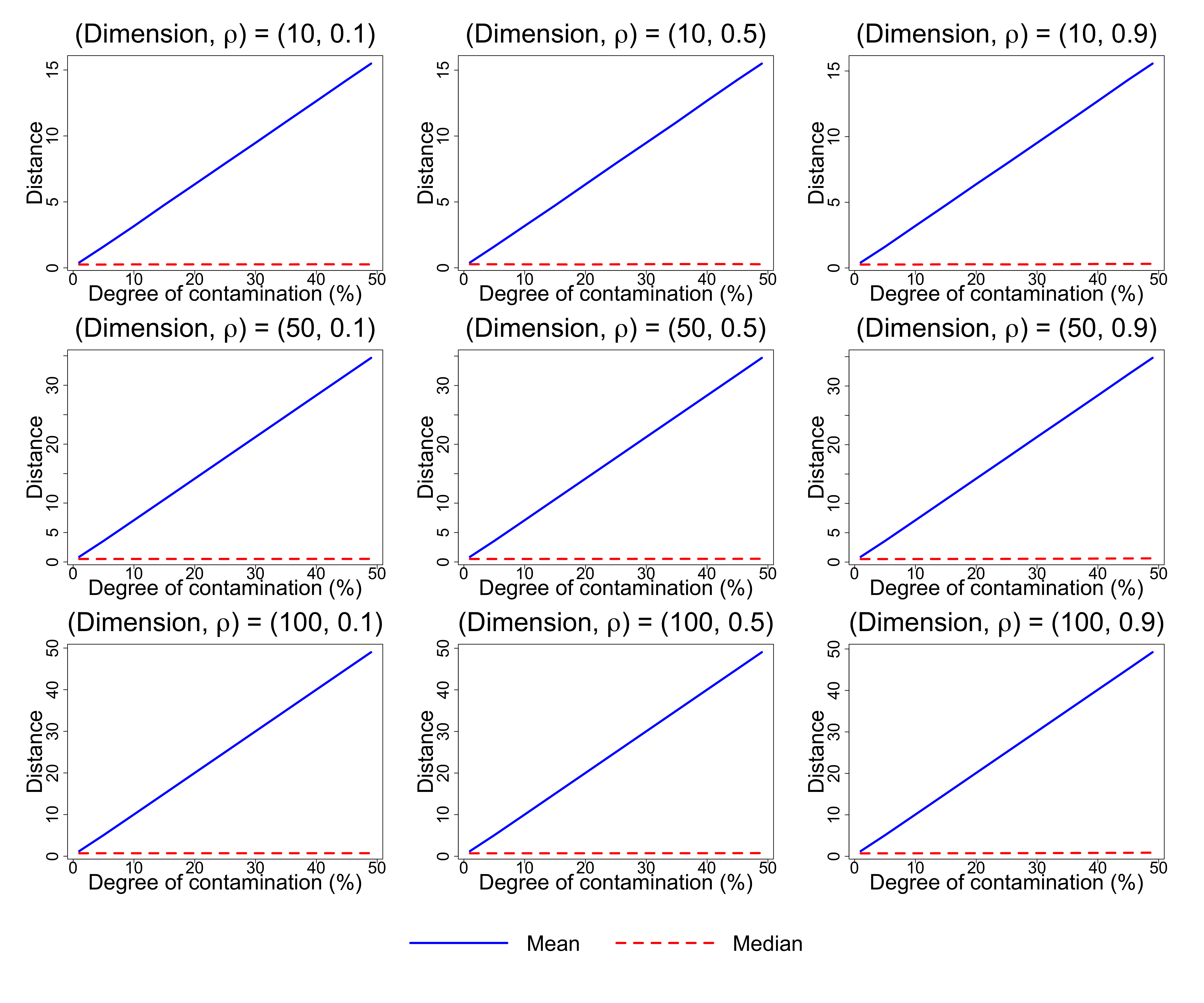}
\caption{Comparison of robustness to contamination between the \Frechet mean and the geometric median of contaminated set of multivarate Gaussian distributions  across different settings of data dimension $d\in \lbrace 10, 50, 100\rbrace$ and autocorrelation structure controlled by $\rho \in \lbrace 0.1, 0.5, 0.9\rbrace$.}
\label{fig:example2}
\end{figure}

Fig. \ref{fig:example2} reports the estimation error as a function of contamination. Across all configurations, the geometric median exhibits markedly better resilience than the \Frechet mean. While the mean degrades steadily under increasing contamination, the median remains robust. The only discernible trend is that the discrepancy grows with dimension,  primarily due to the increased contribution of the mean vectors to the overall Wasserstein distance. These results reinforce the utility of the geometric median for robust estimation on product manifolds, particularly in high-dimensional and contaminated settings.

\section{Conclusion}\label{sec:conclusion}

This work provides a systematic treatment of geometric medians on product manifolds, a setting that naturally arises whenever heterogeneous geometric variables are simultaneously observed. We formulated the median objective in the Riemannian framework and showed that the $\ell_1$ criterion couples the factor manifolds in a fundamentally non-separable manner, which poses challenges to directly utilizing the well-established theory unlike the \Frechet mean. Building on this observation, we developed a general theory establishing existence and uniqueness of the geometric median under curvature and injectivity conditions. In particular, we proved that uniqueness holds globally in Hadamard products and locally under explicit radius constraints, even when one or more components have positive curvature. On the robustness side, we showed that the estimator inherits classical properties: Lipschitz stability under perturbations and an optimal breakdown point of 50\%. These results extend the well-established behavior of the geometric median in Euclidean and single-manifold settings to the product-manifold regime.

On the algorithmic front, we proposed two practical solvers. The first is a Riemannian subgradient method, which is globally convergent under mild assumptions and suitable for general settings. The second is a product-aware Weiszfeld iteration, which achieves local linear convergence when safely away from data singularities. Both methods are designed to update components independently while respecting the coupling structure of the objective. This modularity allows them to leverage existing manifold toolkits and scale effectively to high-dimensional settings. Through examples on the space of univariate and multivariate Gaussians equipped with the Bures-Wasserstein metric, we demonstrated that the geometric median exhibits substantial resilience against contamination, outperforming the \Frechet mean in both accuracy and stability across a range of dimensions and covariance structures.

Despite the breadth of the present analysis, several open directions remain. First, while our theory extends readily to products of more than two manifolds, empirical behavior in higher-order products such as tensor bundles in medical imaging has yet to be fully characterized, particularly under curvature-induced ill-conditioning. Second, our results establish robustness but not yet a complementary statistical theory including consistency and distributional theory. Future work could develop central limit theorems, bootstrap procedures, or inference frameworks for geometric medians in product spaces. Third, the algorithms presented here operate in batch mode. Stochastic or streaming extensions with theoretical guarantees in non-Euclidean settings would be especially valuable for large-scale applications in domains like neuroimaging, robotics, and climate science. Finally, many real-world datasets reside not on strict product manifolds, but on richer structures such as fiber bundles or quotient manifolds. Extending robustness principles and optimization strategies to these more intricate geometries remains an exciting avenue for future research.

\bibliographystyle{dcuky}
\bibliography{references2}

@book{patrangenaru_2016_NonparametricStatisticsManifolds,
	address = {Boca Raton},
	title = {Nonparametric {Statistics} on {Manifolds} and {Their} {Applications} to {Object} {Data} {Analysis}},
	isbn = {978-1-4398-2050-6},
	publisher = {CRC Press, Taylor \& Francis Group},
	author = {Patrangenaru, Victor and Ellingson, Leif},
	year = {2016},
}

@article{takatsu_2011_WassersteinGeometryGaussian,
	title = {Wasserstein geometry of {Gaussian} measures},
	volume = {48},
	number = {4},
	journal = {Osaka Journal of Mathematics},
	author = {Takatsu, Asuka},
	year = {2011},
	pages = {1005 -- 1026},
}

@book{rockafellar_1997_ConvexAnalysis,
	address = {Princeton, NJ},
	edition = {10. print. and 1. paperb. print},
	series = {Princeton {Landmarks} in {Mathematics} and {Physics}},
	title = {Convex {Analysis}},
	isbn = {978-0-691-01586-6 978-0-691-08069-7},
	language = {eng},
	publisher = {Princeton Univ. Press},
	author = {Rockafellar, Ralph Tyrrell},
	year = {1997},
}

@book{lee_2018_IntroductionRiemannianManifolds,
	address = {Cham},
	edition = {2nd ed. 2018},
	series = {Graduate {Texts} in {Mathematics}},
	title = {Introduction to {Riemannian} {Manifolds}},
	volume = {176},
	isbn = {978-3-319-91755-9},
	publisher = {Springer International Publishing : Imprint: Springer},
	author = {Lee, John M.},
	year = {2018},
	doi = {10.1007/978-3-319-91755-9},
}

@book{lee_1997_RiemannianManifoldsIntroduction,
	address = {New York},
	series = {Graduate {Texts} in {Mathematics}},
	title = {Riemannian {Manifolds}: {An} {Introduction} to {Curvature}},
	volume = {176},
	isbn = {978-0-387-98271-7 978-0-387-98322-6},
	shorttitle = {Riemannian manifolds},
	publisher = {Springer},
	author = {Lee, John M.},
	year = {1997},
}

@book{huber_1981_RobustStatistics,
	address = {New York},
	series = {Wiley {Series} in {Probability} and {Statistics}},
	title = {Robust {Statistics}},
	isbn = {978-0-471-41805-4},
	publisher = {Wiley},
	author = {Huber, Peter J.},
	year = {1981},
}

@book{docarmo_1992_RiemannianGeometry,
	address = {Boston},
	series = {Mathematics. {Theory} \& {Applications}},
	title = {Riemannian {Geometry}},
	isbn = {978-0-8176-3490-2 978-3-7643-3490-1},
	language = {eng},
	publisher = {Birkhäuser},
	author = {do Carmo, Manfredo Perdigão},
	year = {1992},
}

@book{petersen_2006_RiemannianGeometry,
	series = {Graduate {Texts} in {Mathematics}},
	title = {Riemannian {Geometry}},
	volume = {171},
	copyright = {http://www.springer.com/tdm},
	isbn = {978-0-387-29246-5},
	url = {http://link.springer.com/10.1007/978-0-387-29403-2},
	language = {en},
	urldate = {2025-05-14},
	publisher = {Springer New York},
	author = {Petersen, Peter},
	year = {2006},
}

@article{frechet_1948_ElementsAleatoiresNature,
	title = {Les éléments aléatoires de nature quelconque dans un espace distancié},
	volume = {10},
	language = {fr},
	number = {4},
	journal = {Annales de l'institut Henri Poincaré},
	author = {Fréchet, Maurice René},
	year = {1948},
	pages = {215--310},
}

@book{cheeger_2008_ComparisonTheoremsRiemannian,
	address = {Providence, R.I},
	series = {{AMS} {Chelsea} {Publishing}},
	title = {Comparison {Theorems} in {Riemannian} {Geometry}},
	isbn = {978-0-8218-4417-5},
	publisher = {AMS Chelsea Publishing},
	author = {Cheeger, Jeff and Ebin, D. G.},
	collaborator = {{American Mathematical Society}},
	year = {2008},
}

@article{you_2025_WassersteinMedianProbability,
	title = {On the {Wasserstein} {Median} of {Probability} {Measures}},
	volume = {34},
	issn = {1061-8600, 1537-2715},
	url = {https://www.tandfonline.com/doi/full/10.1080/10618600.2024.2374580},
	doi = {10.1080/10618600.2024.2374580},
	language = {en},
	number = {1},
	urldate = {2025-05-23},
	journal = {Journal of Computational and Graphical Statistics},
	author = {You, Kisung and Shung, Dennis and Giuffrè, Mauro},
	month = jan,
	year = {2025},
	pages = {253--266},
}

@inproceedings{zhang_2016_FirstorderMethodsGeodesically,
	address = {Columbia University, New York, New York, USA},
	series = {Proceedings of {Machine} {Learning} {Research}},
	title = {First-order {Methods} for {Geodesically} {Convex} {Optimization}},
	volume = {49},
	url = {https://proceedings.mlr.press/v49/zhang16b.html},
	booktitle = {29th {Annual} {Conference} on {Learning} {Theory}},
	publisher = {PMLR},
	author = {Zhang, Hongyi and Sra, Suvrit},
	editor = {Feldman, Vitaly and Rakhlin, Alexander and Shamir, Ohad},
	month = jun,
	year = {2016},
	pages = {1617--1638},
}

@book{mardia_2000_DirectionalStatistics,
	address = {Chichester; New York},
	series = {Wiley series in probability and statistics},
	title = {Directional statistics},
	isbn = {978-0-471-95333-3},
	publisher = {J. Wiley},
	author = {Mardia, K. V. and Jupp, Peter E.},
	year = {2000},
}

@article{beck_2015_WeiszfeldsMethodOld,
	title = {Weiszfeld’s {Method}: {Old} and {New} {Results}},
	volume = {164},
	issn = {0022-3239, 1573-2878},
	shorttitle = {Weiszfeld’s {Method}},
	url = {http://link.springer.com/10.1007/s10957-014-0586-7},
	doi = {10.1007/s10957-014-0586-7},
	language = {en},
	number = {1},
	urldate = {2025-05-24},
	journal = {Journal of Optimization Theory and Applications},
	author = {Beck, Amir and Sabach, Shoham},
	month = jan,
	year = {2015},
	pages = {1--40},
}

@book{absil_2008_OptimizationAlgorithmsMatrix,
	address = {Princeton, NJ},
	title = {Optimization algorithms on matrix manifolds},
	isbn = {978-0-691-13298-3},
	publisher = {Princeton University Press},
	author = {Absil, P.-A. and Mahony, R. and Sepulchre, R.},
	year = {2008},
}

@incollection{selig_2005_LieGroupsLie,
	address = {Dordrecht},
	title = {Lie {Groups} and {Lie} {Algebras} in {Robotics}},
	volume = {136},
	isbn = {978-1-4020-1982-1},
	url = {http://link.springer.com/10.1007/1-4020-2307-3_5},
	language = {en},
	urldate = {2025-05-22},
	booktitle = {Computational {Noncommutative} {Algebra} and {Applications}},
	publisher = {Kluwer Academic Publishers},
	author = {Selig, J. M.},
	editor = {Byrnes, Jim},
	year = {2005},
	doi = {10.1007/1-4020-2307-3_5},
	note = {Series Title: NATO Science Series II: Mathematics, Physics and Chemistry},
	pages = {101--125},
}

@article{ferreira_1998_SubgradientAlgorithmRiemannian,
	title = {Subgradient {Algorithm} on {Riemannian} {Manifolds}},
	volume = {97},
	issn = {0022-3239, 1573-2878},
	url = {http://link.springer.com/10.1023/A:1022675100677},
	doi = {10.1023/A:1022675100677},
	language = {en},
	number = {1},
	urldate = {2025-05-13},
	journal = {Journal of Optimization Theory and Applications},
	author = {Ferreira, O. P. and Oliveira, P. R.},
	month = apr,
	year = {1998},
	pages = {93--104},
}

@article{vardi_2000_MultivariateL1medianAssociated,
	title = {The multivariate {L1}-median and associated data depth},
	volume = {97},
	issn = {0027-8424, 1091-6490},
	url = {http://www.pnas.org/cgi/doi/10.1073/pnas.97.4.1423},
	doi = {10.1073/pnas.97.4.1423},
	language = {en},
	number = {4},
	urldate = {2021-11-11},
	journal = {Proceedings of the National Academy of Sciences},
	author = {Vardi, Y. and Zhang, C.-H.},
	month = feb,
	year = {2000},
	pages = {1423--1426},
}

@article{bickel_2008_RegularizedEstimationLarge,
	title = {Regularized estimation of large covariance matrices},
	volume = {36},
	issn = {0090-5364},
	url = {https://projecteuclid.org/journals/annals-of-statistics/volume-36/issue-1/Regularized-estimation-of-large-covariance-matrices/10.1214/009053607000000758.full},
	doi = {10.1214/009053607000000758},
	number = {1},
	urldate = {2024-03-04},
	journal = {The Annals of Statistics},
	author = {Bickel, Peter J. and Levina, Elizaveta},
	month = feb,
	year = {2008},
}

@article{dryden_2009_NonEuclideanStatisticsCovariance,
	title = {Non-{Euclidean} statistics for covariance matrices, with applications to diffusion tensor imaging},
	volume = {3},
	issn = {1932-6157},
	url = {https://projecteuclid.org/journals/annals-of-applied-statistics/volume-3/issue-3/Non-Euclidean-statistics-for-covariance-matrices-with-applications-to-diffusion/10.1214/09-AOAS249.full},
	doi = {10.1214/09-AOAS249},
	number = {3},
	urldate = {2021-09-27},
	journal = {The Annals of Applied Statistics},
	author = {Dryden, Ian L. and Koloydenko, Alexey and Zhou, Diwei},
	month = sep,
	year = {2009},
}

@article{fletcher_2009_GeometricMedianRiemannian,
	title = {The geometric median on {Riemannian} manifolds with application to robust atlas estimation},
	volume = {45},
	issn = {10538119},
	url = {https://linkinghub.elsevier.com/retrieve/pii/S1053811908012019},
	doi = {10.1016/j.neuroimage.2008.10.052},
	language = {en},
	number = {1},
	urldate = {2021-10-02},
	journal = {NeuroImage},
	author = {Fletcher, P. Thomas and Venkatasubramanian, Suresh and Joshi, Sarang},
	month = mar,
	year = {2009},
	pages = {S143--S152},
}

@article{afsari_2011_Riemannian$L_p$Center,
	title = {Riemannian \$\{{L}\}\_p\$ center of mass: {Existence}, uniqueness, and convexity},
	volume = {139},
	issn = {0002-9939},
	shorttitle = {Riemannian \${L}{\textasciicircum}\{p\}\$ center of mass},
	url = {http://www.ams.org/jourcgi/jour-getitem?pii=S0002-9939-2010-10541-5},
	doi = {10.1090/S0002-9939-2010-10541-5},
	language = {en},
	number = {02},
	urldate = {2021-09-27},
	journal = {Proceedings of the American Mathematical Society},
	author = {Afsari, Bijan},
	month = feb,
	year = {2011},
	pages = {655--655},
}

@book{bhattacharya_2012_NonparametricInferenceManifolds,
	address = {Cambridge},
	title = {Nonparametric {Inference} on {Manifolds}: {With} {Applications} to {Shape} {Spaces}},
	isbn = {978-1-139-09476-4},
	shorttitle = {Nonparametric {Inference} on {Manifolds}},
	url = {http://ebooks.cambridge.org/ref/id/CBO9781139094764},
	urldate = {2021-06-11},
	publisher = {Cambridge University Press},
	author = {Bhattacharya, Abhishek and Bhattacharya, Rabi},
	year = {2012},
	doi = {10.1017/CBO9781139094764},
}

@book{lee_2012_IntroductionSmoothManifolds,
	address = {New York, NY},
	series = {Graduate {Texts} in {Mathematics}},
	title = {Introduction to {Smooth} {Manifolds}},
	volume = {218},
	isbn = {978-1-4419-9981-8 978-1-4419-9982-5},
	url = {https://link.springer.com/10.1007/978-1-4419-9982-5},
	language = {en},
	urldate = {2023-12-08},
	publisher = {Springer New York},
	author = {Lee, John M.},
	year = {2012},
	doi = {10.1007/978-1-4419-9982-5},
}

@book{pennec_2020_RiemannianGeometricStatistics,
	address = {San Diego},
	title = {Riemannian {Geometric} {Statistics} in {Medical} {Image} {Analysis}},
	isbn = {978-0-12-814725-2},
	language = {English},
	publisher = {Academic Press},
	author = {Pennec, Xavier and Sommer, Stefan and Fletcher, Tom},
	year = {2020},
	note = {OCLC: 1128025962},
}

@article{you_2021_RevisitingRiemannianGeometry,
	title = {Re-visiting {Riemannian} geometry of symmetric positive definite matrices for the analysis of functional connectivity},
	volume = {225},
	copyright = {All rights reserved},
	issn = {10538119},
	url = {https://doi.org/10.1016/j.neuroimage.2020.117464},
	doi = {10.1016/j.neuroimage.2020.117464},
	language = {en},
	urldate = {2021-06-09},
	journal = {NeuroImage},
	author = {You, Kisung and Park, Hae-Jeong},
	month = jan,
	year = {2021},
	pages = {117464},
}

\end{document}